
\magnification=\magstep1
\hfuzz=4truept
\nopagenumbers
\font\tif=cmr10 scaled \magstep3

\rightline{PUPT-1481}
\vfil
\centerline{\tif Free energy decreases }\medskip\centerline{\tif
along Wilson renormalization group trajectories}
\vfil
\centerline{{\rm Vipul
Periwal}\footnote{${}^\dagger$}{vipul@puhep1.princeton.edu}}
\bigskip
\baselineskip=18truept
\centerline{Department of Physics}\centerline{Princeton University
}\centerline{Princeton, New Jersey 08544-0708}
\vfil
\par\noindent The free energy is shown to decrease along
Wilson renormalization group trajectories, in a
dimension-independent fashion, for $d>2.$  The argument assumes
the monotonicity of the cutoff function, and
positivity of a spectral representation
of the two point function.  The argument
is valid to all orders in perturbation
theory.
\medskip
\vfil\eject
\def\citr#1{{[{#1}]}}

\def\d{{\hbox{d}}}
\def\be{\beta}
\def\part{\partial}
\def\wil{1}
\def\wz{2}
\def\gross{3}
\def\z{4}
\def\spglcs{5}

The nature of the flow induced by the renormalization group (rg)
on the space of couplings in a field theory
 has been of interest for some time\citr{\wil,\wz,\gross}.
Zamolodchikov's $c$-theorem\citr{\z} rekindled
interest in demonstrating the
dissipative character of the renormalization group.  He demonstrated
that there is a function on the space of renormalizable couplings of a
two-dimensional field theory with the following properties:
\item{(i)} $\d c = 0$ at fixed points,
\item{(ii)} the value of $c$ at these
fixed points is given by the conformal anomaly, and
\item{(iii)} $c$ decreases along rg trajectories, {\it
i.e.}, $\d c(\be) < 0,$ where $\be$ is the vector field associated with
$\be$-functions via $\be \equiv \be^i\part_i.$
\par\noindent Zamolodchikov's argument used the conventional field theory
rg\citr{\spglcs}, with $\be$ satisfying the Callan-Symanzik
equations---the couplings that he needed to consider were only those
associated with marginal and relevant operators.
The $c$ function he found has universal properties, and may therefore
be of physical interest even setting aside
the question of the dissipative character of the field theory rg.

\def\ssc{\scriptscriptstyle}
\def\eps{\epsilon}

\def\wilk{6}
\def\hl{7}
\def\jl{11}
\def\dj{8}
\def\fr{10}
\def\cardy{9}
There is another approach to renormalization due to Wilson\citr{\wilk},
where one studies the flows of all couplings in a cut-off theory.
This set of flows is not universal.  The cut-off is present at all
stages since one has irrelevant operators of arbitrarily high
dimension in the action.  Nevertheless, as can be demonstrated
explicitly, the information contained in the field theory rg can be
extracted from the Wilson rg\citr{\hl}.
I shall demonstrate within the Wilson rg
that the free energy decreases along Wilson rg trajectories,
independent of the dimension of spacetime, for $d>2.$  Thus the Wilson rg
is dissipative, for the class of theories that I am able to consider.
The argument is completely different in character from Zamolodchikov's.
It is unlikely that the free energy will satisfy
either (i) or (ii).  The decrease of the free energy also does not imply
gradient flow, {\it i.e.,} I do not claim that the Wilson rg beta `functions'
are gradients.  I emphasize that the entire framework of the
argument is restricted to all orders in
perturbation theory.  No statements beyond
perturbation theory are suggested or implied in any way.

The field-theoretic rg is not trivially related to the Wilson rg---for some
results in this area, see Refs.~\dj,\hl.
While the intuition for the Zamolodchikov $c$ theorem is usually
couched in terms of `thinning out degrees of freedom', more appropriate
to the Wilson framework, the decrease of the free energy along Wilson rg
trajectories does not
straightforwardly imply the existence
of a $c$-function on the sub-manifold of marginal and relevant couplings.

\def\ha{12}
Attempts
to generalize Zamolodchikov's argument to dimensions larger than two
in simple ways fail because the rotation group is no longer Abelian.
For a review of such attempts, and %
an alternative proof of Zamoldchikov's $c$-theorem, see Ref.~\fr. %
For Wilson rg arguments, see Ref.~\ha.
Cardy\citr{\cardy} suggested that the trace anomaly on a sphere was
the analogue of the central charge in two dimensions,
but the subtractions necessary to define integrals of
two-point functions in continuum field theory render it difficult
to show (iii).

\def\cF{{\cal F}}

I give now a general argument for the decrease of the free energy
for Wilson rg flows.  I will work with Euclidean field theories.
The argument is valid for all theories which
can be regulated by including a cut-off dependence
explicitly in the action.  This is not, perhaps, as large a
class as one wants, since it amounts to theories that (up to cut-off
dependent rescalings of fields) can be regulated by modifying the
kinetic term appropriately.  It applies to all scalar
field theories, and may be extended to theories with fermions
for which one can show the validity of the assumption on the
spectral measure, see below.
\def\dD{\hbox{D}}
\def\ee{\hbox{e}}

The argument requires the derivation
of a new spectral representation---this spectral representation
makes no reference to cutoffs or to the renormalization group, and
hence may be of independent interest.  The derivation is a slight
modification of a derivation given in Ref.~\fr.  Lorentz invariance
implies that the identity operator can be decomposed as
$$1 = \int^\infty_0 \d \mu^2 \int_{M} {\d}^dp \
\delta(p^2+\mu^2)\delta^{(n)}(\hat p-p)\theta(p_0),$$
with $\hat p$ the momentum operator and $p_\alpha=(p_0,{\bf p})$ its Minkowski
eigenvalue.  The subscript $M$ on the integral implies integration over
Minkowski momenta.  For {\it any} monotonically increasing positive
 function, $f$, this can be rewritten as
$$1 = \int^{\infty}_{0} \d (\mu^2f(\mu^2)) \int_{M} {\d}^dp
\ \delta(p^2f(|p^2|)+\mu^2f(\mu^2))\delta^{(d)}(\hat p-p)\theta(p_0).$$
Inserting it into a two-point function (with $x^d<0$) we have
$$\langle\phi(x)\phi(0)\rangle = \int \d m^2
\int_{M}{\d}^dp
\ \exp(ipx)\delta(p^2f(|p^2|)+m^2)\theta(p_0)\langle0|\phi(0)
\delta^{(d)}(\hat p-p)\phi(0)|0\rangle .$$
The matrix element is independent of $x,$ and is just a function
of $p^2.$  The integral over $p$ can be done, giving
$$\int_{M}{\d}^dp \ee^{ipx}\delta(p^2f(|p^2|)+m^2)\theta(p_0)=
\int {\d}^{d-1}{\bf p} \ee^{i{\bf p}\cdot {\bf x}}\int_0^\infty \d p_0
\ee^{p_0x^d}\delta(({\bf p}^2-p_0^2)f(|{\bf p}^2-p_0^2|)+m^2).$$
Finally, since $x^d<0$ by assumption, the integral over $p_0$ with a
$\delta$ function can be rewritten as
$$\int {\d}^{d-1}{\bf p}\ \exp(i{\bf p}\cdot {\bf x})\int_{-\infty}^\infty
{{\d p_d}\over{2\pi}}\exp(ip_dx^d) {{\theta(-x^d)}\over{
({\bf p}^2+p_d^2)f({\bf p}^2+p_d^2)+m^2}}.$$
Add the term with $x^d>0$ to get the propagator in Euclidean space.
Thus, putting terms together, one finds the spectral representation
$$\langle\phi(p)\phi(-p)\rangle = \int^{\infty}_{0}
 \d m^2\rho(m^2) {1\over {p^2f(p^2)+m^2}},$$
with
$$\rho(m^2) = (2\pi)^{d-1}\langle0|\phi(x=0)\delta(\hat p -p)\phi(x=0)|0
\rangle|_{p^2f(|p^2|)=-m^2}.$$
Note that
nowhere in this derivation is {\it any} use made of the cutoff nature of
the theory.  { Based on this fact}, I assume that $\rho$ is positive,
since this is a two-point function with the insertion of a positive operator
delta function, and the two-point function should be positive even
in the cutoff theory.

Let $Z_{\ssc 0} \equiv \int \prod \dD \phi \exp({-S_{int}(\phi)-{1\over
2}\int \phi(-\part^2)\phi})$ be
a formal partition function, expressed as a functional integral over some
set of fields, $\phi.$
A proper definition  of $Z_{\ssc 0}$ requires
a cutoff.  Let $K(x)$ be a differentiable monotone function
such that
$${K(x) = 1\qquad {\rm for}\  x<1, \qquad\hbox{and}\qquad
K(x) \downarrow 0\qquad {\rm for}\ x\gg 1.}$$
I assume that $K(x)$ goes to zero faster than any negative power of $x$
for $x$ large.  Suppose that
$$Z_\eps[J] \equiv \int{\dD\phi\over{\det^{1\over 2}K(-\part^2/\eps^2)
(-\part^{-2})}}  \exp\left(-S_{int}(\phi) -{1\over
2}\int \phi(-\part^2 ) K^{-1}(-\part^2 /\eps^2) \phi - \int J\phi\right)
$$
is well-defined for any $\eps<\infty.$  As a formal limit,
$\lim_{\eps\uparrow\infty}Z_\eps[J=0] = Z_{\ssc 0}.$
One can be more general, as will be apparent from the
reasoning below.  $J$ is a source restricted to vanish for momenta
of order $\eps$ or larger.  The Wilson rg flow is defined by demanding the
invariance of $Z_\eps[J],$ for such $J.$  It is easy to derive such a
flow, not just for the low momentum correlation functions, but for
$Z_\eps[J]$ itself---the normalization of the measure is important for
this.

The idea behind what follows is simple: decreasing $\eps$ is equivalent
to gradually integrating out modes with large momenta.
The change in the free energy due to this explicit change in the
cutoff is therefore of definite sign, due to the monotonicity of $K.$
This is true {\it even} taking into account the
normalization of the functional measure,
{\it assuming} the positivity of the spectral representation of the
two point function.  On the other hand, by the definition of
the Wilson rg, the free energy is left invariant under changes in
$\eps,$ provided the couplings in $S_{int}
\equiv \sum g^{\ssc A}(\eps) \int \Phi_{\ssc A}$
are appropriately adjusted, so
the change in the free energy upon just changing the couplings must
be of the opposite sign as the change due to the explicit presence of
the cutoff.  $\{\Phi_{\ssc A}\}$ is
a complete set of scalar operators, including the identity.

\def\vv{{1\over \hbox{Vol.}}}
Define $\cF\equiv- \ln Z_\eps[0]/{\rm Vol.},$ the free energy density.
All expressions with $1/{\rm Vol.}$ factors should be understood
as implying an infinite volume limit.  We have
$$\eqalign{0 = \eps{\d\over\d\eps}\cF =
 &\vv\Big\langle\eps{\part\over\part\eps}S_{int}\Big\rangle
\cr & +\ \vv \Big\langle {1\over 2}
\int \phi \eps{\part\over\part\eps}K^{-1}(-\part^2/\eps^2)(-\part^2)\phi
\Big\rangle -{1\over 2} \int K\eps{\part\over\part\eps}K^{-1} .} $$
This amounts to
$$ \vv\Big\langle\eps{\part\over\part\eps}S_{int}\Big\rangle
+\vv \int{1\over 2}
\eps{\part\over\part\eps}K^{-1}(p^2/\eps^2)p^2
\left(\langle\phi\phi\rangle - {K(p^2/\eps^2)\over p^2}
\right)=0.$$
I now wish to argue that the second term is positive on general
grounds.

By picking $f(x)=  K(x/\eps^2)^{-1} ,$ the spectral representation becomes
$$
\langle\phi(p)\phi(-p)\rangle = \int_0^\infty \d m^2 \rho(m^2)
{K(p^2/\eps^2)\over p^2+K(p^2/\eps^2)m^2},$$
with
$$ \rho(m^2) \equiv
(2\pi)^{d-1}
\langle 0|\phi(x=0)\delta(\hat p -p)\phi(x=0)|0
\rangle|_{p^2+K(|p^2|/\eps^2)m^2=0} .$$
An implicit assumption here is the existence of
`vacuum' boundary conditions
for the functional integration---due to the higher derivative form
of the cutoff theory, a Hamiltonian definition is problematic.

{}From the spectral representation, we note
$$\lim_{p^2\uparrow\infty}p^2K^{-1}(p^2/\eps^2)\langle\phi(p)\phi(-p)\rangle
= \int_0^\infty \d m^2 \rho(m^2).$$
In cutoff perturbation theory, with the assumed form of $K,$ we therefore have
$$\int \d m^2 \rho(m^2)=1.$$
This is valid to all orders in perturbation theory.
This normalization of the spectral measure in the cutoff theory, and
the positivity of the spectral measure, are the
crucial assumptions in the present discussion.
The normalization assumption is not true for interacting renormalized
field theories, hence the present argument does not admit any na\"\i ve
generalization to the field theoretic rg.

It now follows that
$$\vv \int{1\over 2}
\eps{\part\over\part\eps}K^{-1}(p^2/\eps^2)p^2
\left(\langle\phi\phi\rangle - {K(p^2/\eps^2)\over p^2}
\right)>0,$$
since
$${K(p^2/\eps^2)\over p^2} \ge
{K(p^2/\eps^2)\over p^2+K(p^2/\eps^2)m^2}$$
for $m^2\ge 0,$ and $K$ is monotone.
This quantity is certainly finite to all orders in perturbation theory, since
we have normalized the measure appropriately.
This finally implies
$$\vv\Big\langle\eps{\part\over\part\eps}S_{int}\Big\rangle=
\sum \eps{\part\over\part\eps}g^{\ssc A}(\eps)\vv \Big\langle\int
\Phi_{\ssc A}\Big\rangle\equiv \sum \beta_{\ssc W}^{\ssc A}
\part_{\ssc A} \cF <0.$$
This is precisely what we wished to show, since
we have shown that $\cF$ decreases along Wilson rg
trajectories, defined by the vector field $\be_{\ssc W}$,
at every fixed $\eps.$

Some caveats: (1) If the Wilson rg flow leads to theories where the positivity
assumption
no longer holds, this argument breaks down. Such theories would not appear to
make sense as Euclidean field theories.  (2) This argument does not
extend in any straightforward way to gauge theories, because the only
gauge-invariant `physical' regulator known is the lattice, where the
separation of kinetic and interaction terms in the action is unnatural.
(3) This argument only holds to all orders in perturbation theory.

It may be possible to use this decrease of $\cF$ to derive a $c$ theorem
for the field theoretic rg, even though the intermediate steps cannot be
translated.  The necessary ingredients are (a) to control the manner in
which the Wilson rg flow converges to the submanifold of relevant and
marginal couplings, and (b) to control a change of variables from the
couplings that occur in the cutoff action to the normalization conditions
that describe this relevant/marginal coupling submanifold.
\def\ha{12}
\bigskip
Acknowledgements:  I am grateful to R.C. Myers for helpful
conversations.  I would also like to thank J. Cardy, D. Friedan,
D. Gross and M. L\"assig for useful discussions and comments.
This work was supported in part by NSF Grant No. PHY90-21984.
\hfuzz=2pt
\bigskip
\centerline{References}
\bigskip
\item{\wil.} K. Wilson, in {\it Collective properties of physical
systems}, (B. Lundqvist and S. Lundqvist, eds.),
Academic Press, New York, 1973

\item{\wz.} D.J. Wallace and R.K.P. Zia, {\sl Ann. Phys.} {\bf 92} (1975)
142

\item{\gross.} D.J. Gross, in {\it Field theory and critical phenomena},
(R. Balian and J. Zinn-Justin, eds.), North-Holland, Amsterdam, 1975

\item{\z.} A.B. Zamolodchikov, {\sl JETP Lett.} {\bf 43} (1986) 730

\item{\spglcs.} E.C.G. Stueckelberg and A. Petermann, {\sl Helv. Phys.
Acta} {\bf 26} (1953) 499; M. Gell-Mann and F.E. Low, {\sl Phys. Rev.}
{\bf 95} (1954) 1300; C.G. Callan, {\sl Phys. Rev.} {\bf D2} (1970)
1541; K. Symanzik, {\sl Comm. Math. Phys.} {\bf 23} (1971) 227

\item{\wilk.} K.G. Wilson and J.B. Kogut, {\sl Phys. Rep.} {\bf 12C}
(1974) 75; F. Wegner, in {\it Phase transitions
and critical phenomena}, vol. 6, (C. Domb and M. Green, eds.), Academic
Press, New York, 1976; J. Polchinski, {\sl Nucl. Phys.} {\bf B231}
(1984) 269

\item{\hl.} J. Hughes and J. Liu, {\sl Nucl. Phys.} {\bf B307} (1988)
183

\item{\dj.} C. di Castro and G. Jona-Lasinio, in {\it Phase transitions
and critical phenomena}, vol. 6, (C. Domb and M. Green, eds.), Academic
Press, New York, 1976

\item{\cardy.} J. Cardy, {\sl Phys. Lett.} {\bf 215B} (1988) 749

\item{\fr.} A. Cappelli, D. Friedan and J. Latorre, {\sl Nucl. Phys.}
{\bf B352} (1991) 616

\item{\jl.} G. Jona-Lasinio, in {\it Collective properties of physical
systems}, (B. Lundqvist and S. Lundquist, eds.),
Academic Press, New York, 1973

\item{\ha.} P. Haagensen, Y. Kubyshin, J. Latorre and E. Moreno, `Gradient
flows from an approximation to the exact renormalization group',
Barcelona preprint (1993), and references therein.
\end

\end